%% file: hanl1b.tex
\documentclass[twocolumn,showpacs,preprintnumbers,amsmath,amssymb]{revtex4}

\begin{document}

\title{Stability and correlations in dilute two-dimensional boson systems}

\author{Han Guangze} 

\affiliation{Department of Applied Physics, South China University of
  Technology, Guangzhou 510640, China}

\author{O.~S{\o}rensen} \email{oles@phys.au.dk}

\author{A.~S.~Jensen} 

\author{D.~V.~Fedorov}

\affiliation{ Department of Physics and Astronomy, University of
  Aarhus, DK-8000 Aarhus C, Denmark }

\date{\today}

\begin{abstract}
  The hyperspherical adiabatic expansion method is used to describe
  correlations in a symmetric boson system rigorously confined to two
  spatial dimensions.  The hyperangular eigenvalue equation turns out
  to be almost independent of the hyperradius, whereas the solutions
  are strongly varying with the strength of the attractive two-body
  potentials.  Instability is encountered in hyperangular,
  hyperradial, and mean-field equations for almost identical strengths
  inversely proportional to the particle number.  The derived
  conditions for stability are similar to mean-field conditions and
  closely related to the possible occurrence of the Thomas and Efimov
  effects.  Renormalization in mean-field calculations for two spatial
  dimensions is probably not needed.
\end{abstract}


\maketitle

\paragraph*{Introduction.}

Lower dimensions than three are necessary in several branches of
physics, for instance surface physics, semiconductor physics,
artificial atoms, and quantum dots.  The advanced tools of
well-controlled external fields, used in atomic and molecular physics
to manipulate the effective interactions, allow confinement of the
systems to lower dimensions \cite{gor01b,ryc03}.  The experimental
investigations employ continuous variation of the dimension by use of
tunable deformed external fields.

The basic properties vary dramatically with the dimensionality of the
system, as highlighted in two dimensions ($2D$) where the centrifugal
$s$-wave barrier is negative for two particles and even an
infinitesimally small attraction provides a bound state
\cite{nie01,khu02}.  The stability is strongly dependent on the
deformation or the effective dimension of the confining potential
\cite{rob01,gam01,adh01c}.  The simplest $N$-body structures are the
Bose-Einstein condensates for identical bosons.  These systems are
dilute, weakly interacting, and well described by mean-field models
\cite{pet01,pit03} with an interaction strength adjusted to reproduce
the low-energy scattering cross section in the Born approximation.
The interaction in $2D$ is then obtained from short- or zero-range
three-dimensional ($3D$) potentials restricted to the mean-field
Hilbert space \cite{lee02,raj03}.

Inclusion of correlations in the wave function prohibits this
renormalization.  Instead a finite short-range potential with the
correct scattering length should be used.  If the correlations are
appropriately accounted for, the large-distance behavior must come out
correctly with the realistic interaction \cite{sor03a,sor03b}.  The
consequences for lower dimensions are not yet investigated.  The huge
difference between two- and three-body properties in two and three
dimensions is most likely more pronounced for $N$-body systems.
Experimental results from varying dimensionality are easier to
interpret if the limits are known.

The purpose of this letter is to (i) formulate a framework for
investigations of correlated structures under strict $2D$ confinement
for $N$ interacting identical bosons in an external harmonic field,
(ii) derive stability conditions in terms of particle number and
two-body interaction properties, (iii) extract the basic features of
the solutions, and (iv) provide a conceptual link between the
successfully renormalized mean-field models and the correlated
solutions with the bare (effective) interaction.

\paragraph*{Theoretical method.}

We shall briefly sketch the method to establish the notation and the
pertinent formulae for two dimensions.  We follow the derivation for
three dimensions given in \cite{sor02b}.  The $N$ identical bosons
have masses $m$ and coordinates $\vec r_i$.  We use the hyperspherical
adiabatic expansion method where the only length coordinate is the
hyperradius $\rho$ defined by
\begin{equation} 
  \rho^2 \equiv \frac1N\sum_{i<j}^Nr_{ij}^2=\sum_{i=1}^Nr_i^2-NR^2
  \label{e3}
  \;,
\end{equation}
where $\vec r_{ij} = \vec r_i - \vec r_j$ and $\vec R = \sum_i\vec
r_i/N$ is the center-of-mass coordinate.  The remaining $2N-3$
relative coordinates in two dimensions are angles where we define
$\alpha_{ij}$ related to the size of $\vec r_{ij}$ by $r_{ij} \equiv
\sqrt{2} \rho \sin \alpha_{ij}$.  If permitted in the context, we
shall omit the indices $ij$.

The center of mass separates out and we only need to deal with
relative coordinates.  The related volume element is $\rho^{2N-3} {\rm
  d}\rho \sin \alpha \cos^{2N-5} \alpha {\rm d} \alpha {\rm d}
\vartheta {\rm d} \Omega_{N-2}$, where $\alpha = \alpha_{ij}$,
$\vartheta$ describes the direction of $\vec r_{ij}$ and
$\Omega_{N-2}$ denotes the remaining angular part of the volume
element corresponding to the last $N-2$ relative vector coordinates.

An external harmonic potential $m\omega^2 \sum_i r_i^2/2$ of angular
frequency $\omega$ is by use of eq.~(\ref{e3}) divided into a
center-of-mass part and a hyperadial part.  The relative Hamiltonian
is then separated into a hyperradial part and a hyperangular part,
$\hat h_\Omega$, i.e.
\begin{eqnarray}
  \hat H 
  =   \frac{\hbar^2}{2m} \Big(
  -\frac{1}{\rho^{2N-3}} \frac{\partial}{\partial\rho} \rho^{2N-3}
  \frac{\partial}{\partial\rho} +\frac{\rho^2}{b_t^4}
  +\frac{\hat h_\Omega}{\rho^2}  \Big) 
  \;, 
\end{eqnarray}
\begin{eqnarray}
  &&
  \hat  h_\Omega
  = \hat\Pi^2+  D_{angle}
  +\frac{2m\rho^2}{\hbar^2}\sum_{i<j}^N \hat V(r_{ij})
  \;,
  \\
  &&
  \hat\Pi^2
  = -\frac{\partial^2}{\partial\alpha^2}+
  \frac{2N-6 - (2N-4)\cos 2\alpha}{\sin 2 \alpha}
  \frac{\partial}{\partial\alpha}
  \label{e13}
  \;,
\end{eqnarray}
where all differential angular dependence, except $\alpha$, is
collected in $D_{angle}$.  The trap length $b_t$ is given by $b_t^2
\equiv \hbar/(m\omega)$.  The two-body interaction $\hat V_{ij}$ is of
short range, e.g. a Gaussian $V_0 \exp(-r_{ij}^2/b^2)$ or a square
well $V_0\Theta(r_{ij}<b)$, where $\Theta$ is the truth function.

The relative wave function $\Psi(\rho,\Omega)$ obeys the Schr\"odinger
equation 
\begin{eqnarray}
  \hat H\Psi(\rho,\Omega)
  =
  E\Psi(\rho,\Omega)
  \;,
\end{eqnarray}
where $E$ is the energy.  We write $\Psi$ as an adiabatic expansion
\cite{nie01,sor02b} where the first term is
\begin{eqnarray}   
  \Psi(\rho,\Omega)
  =
  \rho^{-(2N-3)/2} f(\rho) \Phi(\rho,\Omega)
  \label{e42}
  \;,
\end{eqnarray}
with the hyperradial volume element explicitly extracted.  The angular
wave function $\Phi(\rho,\Omega)$ is for fixed $\rho$ an eigenfunction
of $\hat h_\Omega$ with the eigenvalue $\lambda(\rho)$, i.e.
\begin{eqnarray}
  \hat h_\Omega\Phi(\rho,\Omega)
  =
  \lambda(\rho)\Phi(\rho,\Omega)
  \;.
  \label{e25}
\end{eqnarray}
The corresponding radial equation is then
\begin{eqnarray}
  &&
  \Big(-\frac{\hbar^2}{2m}\frac{d^2}{d\rho^2} + U(\rho) - E\Big)
  f(\rho)
  =
  0
  \;,
  \label{e32}
  \\
  &&
  \frac{2mU(\rho)}{\hbar^2}
  =
  \frac{\lambda (\rho)}{\rho^2}+
  \frac{(2N-3)(2N-5)}{4\rho^2}+
  \frac{\rho^2}{b_t^4}
  \;,
  \label{e35}
\end{eqnarray}
where the adiabatic potential $U$ is a function of the hyperradius
consisting of three terms, i.e.~the angular average $\lambda$ of the
interactions and kinetic energies, the generalized centrifugal
barrier, and the external field.

For large particle distances only relative $s$ waves contribute.  With
a Faddeev decomposition of the angular wave function only the
dependence on distance $\alpha_{ij}$ is left, i.e.
\begin{eqnarray} \label{e30}
  \Phi(\rho,\Omega) =
  \sum_{i<j}^N   \tilde \phi(\rho,\alpha_{ij}) \equiv
  \sum_{i<j}^N  \frac{\phi(\rho,\alpha_{ij})}{\sin^{\frac12}
    \alpha_{ij} \cos^{N-\frac52}\alpha_{ij}}
  \;,
\end{eqnarray}
where we again explicitly extracted the square root of the volume
element.

The integro-differential equation for $\phi(\rho,\alpha)$ is obtained
from eq.~(\ref{e25}) by integrating over all other angles than
$\alpha_{12}=\alpha$, denoted by $\tau$ \cite{sor02b}, i.e.
\begin{eqnarray} 
  \bigg(
  -\frac{\partial^2}{\partial\alpha^2} 
  +  v_c + v +  v_1   + v_2  - \lambda
  \bigg) 
  \phi(\rho,\alpha) 
  =\int d \tau \; G
  \label{e40}
  \;,\;
\end{eqnarray}
where $G=G(\tau,\alpha)$ is linear in both $\phi$ and $V$
\cite{sor02b}, and the reduced potentials are given by
\begin{eqnarray} 
  v_c \equiv  
  \frac{(2N-5)(2N-7)}{4}\tan^2\alpha  
  -\frac{\cot^2 \alpha}{4} 
  -\frac{4N-9}{2} 
  \label{e59} 
  \;,\;
\end{eqnarray}
\begin{eqnarray} \label{e50}
  &&
  v(\alpha) =  4 a_B\big(\frac{\rho}{b}\big)^2 \exp(-2 (\rho \sin \alpha /b)^2)
  \;,\\
  &&
  v_1(\alpha) \equiv 2 a_B \big(\frac{\rho}{b}\big)^2 (N-3)^2(N-2)  
  \nonumber\\
  &&
  \qquad
  \times  \int {\rm d}x x^{N-4}   \exp{[2(x-1)(\frac{\rho}{b}\cos \alpha)^2]}
  \label{e56}
  \\
  &&
  \qquad
  \approx \frac{(N-3)^2(N-2)a_B}{\cos^2 \alpha}
  \label{e51}
  \;,\\
  &&
  v_2(\alpha) \approx \frac{4}{3} v_1(\alpha) (1-\frac{1}{3}
  \tan^2 \alpha)^{N-4} \Theta(\alpha < \pi/3) 
  \label{e52}
  \;,\quad
  \\
  &&
  a_B \equiv  \frac{m}{2 \pi \hbar^2} \int V(r) d\vartheta r {\rm d}r
  = \frac{m V_0 b^2}{2 \hbar^2} \label{e54}
  \;.
\end{eqnarray}
The last expression for $a_B$ is valid both for Gaussian and
square-well potentials $V$.  The approximations for $v_1$ and $v_2$
are very accurate for $\rho \cos \alpha \gg b$.  For $\alpha = \pi/2$
we get exactly $v_1(\pi/2)= 2(N-3)(N-2)a_B (\rho/b)^2$, and $v_2$ is
very small.

For a short-range interaction the right-hand side of eq.~(\ref{e40})
is independent of $\rho$ as well as $v_1$ and $v_2$ when $\alpha$ is
not too close to $\pi/2$.  The only $\rho$ dependence is then through
$v$, which approaches a zero-range interaction in $\alpha$ as $\rho$
increases. The eigenvalue $\lambda(\rho)$ is therefore expected to be
constant in large ranges of $\rho$. These features are unique for two
dimensions.

\paragraph*{Stability conditions.}

For comparison we first consider the mean-field approximation in two
dimensions.  This was investigated in details in \cite{zyl02} with a
potential derived from a three-dimensional zero-range potential by
\cite{pet00}. Neglecting the logarithmic energy dependent term in
\cite{zyl02,pet00} only a two-dimensional zero-range potential remains
precisely as the delta-function limit of our Gaussian short-range
potential. In general, for an interaction of short range, i.e.~small
$b$, the differential equation is in dimensionless quantities given by
\begin{eqnarray}
  &&
  \bigg(
  -\frac{\partial^2}{\partial x^2}  -\frac{1}{x}\frac{\partial}{\partial x}
  + \Big(\frac{b}{b_t}\Big)^4 x^2 +
  \label{e53}
  \\ \nonumber
  &&
  \qquad
  2 a_B (N-1) |f_m(x)|^2 -\epsilon \bigg) f_m(x) = 0
  \;,
\end{eqnarray}
where $x\equiv r/b$ and $\epsilon \equiv 2m E_m b^2/\hbar^2$ are
measures of the single-particle mean-field coordinate $r$ and energy
$E_m$. The radial wave function $f_m$ is approximated by a Gaussian,
i.e.~$f_m = \exp[-x^2/(2 d^2)]/(\sqrt{\pi}d)$ renormalized as $\int
d\vartheta x dx |f_m(x)|^2 = 1$ as in \cite{zyl02}.  The corresponding
energy per particle $\epsilon$ is then as a function of $d$ given by
\begin{eqnarray} 
  \epsilon 
  =
  \Big(\frac{b}{b_t}\Big)^4 d^2 + \frac{1}{d^2}\big[1+a_B (N-1)\big]
  \label{e55}
  \;,
\end{eqnarray}
which only has a minimum when 
\begin{eqnarray} \label{e77}
  a_B(N-1) > - 1 \; .
\end{eqnarray}
Then the energy and the Gaussian width are
\begin{eqnarray} \label{e60}
  &&
  d = \frac {b_t}{b} \big[1+ (N-1) a_B\big]^{1/4}
  \;,
  \\
  &&
  \epsilon  =  2 \Big(\frac{b}{b_t}\Big)^2 \sqrt {1+ (N-1) a_B}
  \;.
\end{eqnarray}
These results coincide for $N-1 \approx N$ and $a_B = \tilde g$ with
those derived in \cite{zyl02}.  If the interaction strength in an
experiment suddenly is changed from repulsive to an attractive value
$a_B$, the new state of the $N$-body system can only be stable for
particle numbers smaller than the critical value $N_c = 1-1/a_B$.  For
$N>N_c$ the motion is towards larger densities and either a total
collapse or a reduction of particles in the gas caused by molecular
recombination. 

Thus, only repulsive or very weakly attractive potentials provide
stable mean-field solutions. The basic reason is that even
infinitesimally small attractions bind two particles in two
dimensions. The corresponding two-body Schr\"{o}dinger equation with
wave function $f_2$ and energy $E_2$ is
\begin{eqnarray}
  \bigg[ -\frac{\partial^2}{\partial x^2}  -\frac{1}{x}\frac{\partial}{\partial x}
  + 2 a_B \Big(\frac{V(x) - E_2}{V_0}\Big)\bigg] f_2(x) = 0 
  \label{e65}  
  \;,
\end{eqnarray}
which for a weakly attractive interaction has the bound-state energy
$E_2= - 4 \hbar^2 /(m b^2) \exp(2/a_B-2\gamma)$ and the mean-square
radius $\langle r^2 \rangle = 2 \hbar^2 / (3 m |E_2|)$, where $\gamma$
is Euler's constant, see \cite{nie01}.

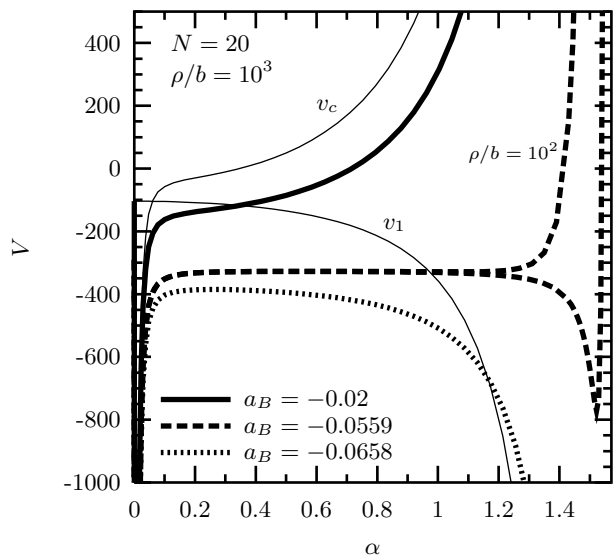
\begin{figure}[htb] 
  \input{hanl1fig1.tex}
  \caption[] {The centrifugal terms $v_c$ and $v_1$ from
    eqs.~(\ref{e59}) and (\ref{e56}) as function of $\alpha$.  The
    thick lines show $v_c + v_1$ for $N=20$, various values of $a_B$,
    and $\rho/b=10^3$ unless otherwise indicated.  The thin, solid
    lines show the contributions from $v_1$ and $v_c$ for
    $a_B=-0.02$. }
  \label{fig1}
\end{figure}

The radial potential in eq.~(\ref{e35}) depends crucially on $\lambda$
determined from eq.~(\ref{e40}) which in turn is dominated by the
terms $v_c$ and $v_1$ shown in fig.~\ref{fig1}.  When $\alpha$
approaches $\pi/2$ the approximation in eq.~(\ref{e51}) and $v_c$
eq.~(\ref{e59}) both diverge as $\cos^{-2}\alpha \approx
(\pi/2-\alpha)^{-2}$.  Thus, when $a_B < -(N-5/2) (N-7/2) /
[(N-2)(N-3)^2] \approx -1/(N-2)$ an attractive pocket inevitably
appears for large $\rho$.  This divergence at the point $\alpha =
\pi/2$ disappears when the exact expression in eq.~(\ref{e56}) is used
for $v_1$. Then the lowest eigenvalue $\lambda$ would be finite and
proportional to $\rho^{-2}$.  These solutions correspond to many
particles close together which probably violates our assumption of
$s$-wave dominance in the wave function. To avoid this divergence at
large $\rho$ the strength of the attraction must be limited by $a_B
(N-2) > -1$, which is almost identical to eq.~(\ref{e77}).  Thus,
remarkably enough the mean-field stability condition is precisely also
obtained from the angular potential when $N-2 \approx N$.

For small $\alpha$ the angular centrifugal term $v_c$ from
eq.~(\ref{e59}) diverges as $-\alpha^2/4$. This is the limit
rigorously separating attractions leading to either no bound states or
infinitly many bound states of Thomas or Efimov character
\cite{nie01}.  Thus, a small two-body attraction $v$ is for large
$\rho$ sufficient to bind a state in the pocket at small $\alpha$.
This is seen by substituting $x\equiv \sqrt{2} \rho \alpha / b$ in
eq.~(\ref{e40}) when only the potential $v$ and $v_c$ are included.
The energy and the mean-square radius of the solution becomes
$\lambda(\rho) \rightarrow 2m\rho^2 E_2 /\hbar^2$ and $\langle
\alpha^2\rangle \rightarrow 2 /(3|\lambda|) = 1/12 (b/\rho)^2
\exp(2\gamma-2/a_B)$.  If the size in $\alpha$ space has to be smaller
than unity, $\rho/b$ must exceed $\exp(-1/a_B)$ which is huge when
$-1/N < a_B <0$, i.e.~the interaction is attractive but allows
physical solutions in agreement with eq.~(\ref{e77}).  Thus, the
diverging $\lambda$ corresponding to the bound two-body state is never
encountered because either the interaction is too attractive leading
to solutions in the pocket at large $\alpha$, or the interaction is
too weak to bind at small $\alpha$ for $\rho$ values less than the
trap length. Therefore diatomic recombination is unlikely.

In any case the angular potential provides two types of minima at
small and large $\alpha$-values, respectively. The large
$\alpha$-behavior corresponds to the stability condition obtained by
mean-field calculations. The structure attempts to maximize the
two-body attractions by a rather similar distance between all
particles confined to the volume allowed by the given hyperradius,
i.e.~two particles are far apart but all others are correspondingly
close.

The small $\alpha$-behavior turns out to be unimportant for stability,
because the extremely weak binding requires a spatially extended wave
function attempting to reach beyond the confining boundaries of the
trap.  A huge trap would in principle allow binding by this pocket,
which corresponds to a structure with one two-body bound state and
consequently this does not resemble a condensed state. Such a
structure would not be found by mean-field calculations although the
two-body bound state always is present for overall attraction in two
dimensions.

The deeper-lying reason for not populating this two-body bound state
seems to be related to the choice of strength for a two-dimensional
zero-range interaction. The $2D$ scaling property of a finite range
(Gaussian) interaction matches with maintaining fixed values of both
the scattering length and the two-body bound state energy.  Thus, both
mean-field zero-range and finite-range correlation calculations can
use the strength determined to reproduce a given value of the
scattering length in the Born approximmation.  This is not possible in
three dimensions, see the renormalization paragraph below.

\paragraph*{Radial solutions.}

To solve the hyperradial eq.~(\ref{e35}) we need $\lambda$.  For
$\rho=0$ only $v_c$ is present in eq.~(\ref{e40}).  The free angular
solutions are
\begin{eqnarray} \label{e45}
  \tilde \phi_{\nu} = P^{(0,N-3)}_{\nu} (\cos 2 \alpha)
  \;,\quad
  \lambda_{\nu}^{(f)} = 4\nu(\nu + N-2) 
  \;,
\end{eqnarray}
where $P^{(0,N-3)}_{\nu}$ is the Jacobi function \cite{nie01} and $\nu
=0,1,2,3 ...$ is a non-negative integer.  For very small $\rho$
perturbation then gives $\lambda \approx \lambda_{\nu}^{(f)} + 2 a_B
N(N-1) (\rho/b)^2$.

Increasing $\rho$ for a repulsive or weakly attractive potential
leaves approximately the free wave function, but the energies are
shifted from the contribution of the interaction. In first-order
perturbation with the free wave functions the eigenvalues are denoted
$\lambda^{(\delta)}_{\nu}$, where the lowest for $\rho \gg b$ is found
to be
\begin{eqnarray}
  \lambda^{(\delta)}_{\nu=0} = a_B N(N-1)(N-2)
  \;.
\end{eqnarray}
When the attraction is stronger, both wave function and energy
change. However, the only $\rho$ dependence in the angular equation is
in $v$ and in $v_1$ when $\alpha \approx \pi/2$.  Still, the
eigenvalues $\lambda$ are essentially $\rho$ independent when $b \ll
\rho \ll \exp(-1/a_B)$, but perhaps lower than $\lambda_{\nu}$.  The
eigenvalues must be obtained by numerical calculations.

\begin{figure}[htb]
  \input{hanl1fig2.tex}
  \caption[] {The lowest angular eigenvalues as functions of $\rho$
    for different values of $a_B$ and $N$. The thick, solid line is
    $\lambda_0$ for $N=20$ and $a_B=-0.02$.  The parameters for the
    other curves are given directly on the figure.}
  \label{fig2}
\end{figure}
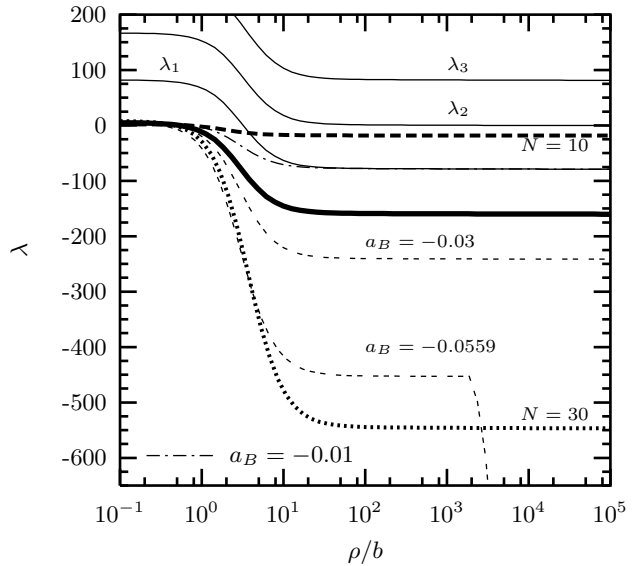

The behavior of the angular eigenvalues in the interesting range of
$\rho$ values are illustrated in fig.~\ref{fig2}.  They can be
parametrized rather well by
\begin{eqnarray}
  &&
  \lambda (\rho) \approx \frac{\lambda_{\nu}^{(f)} - \lambda^{(\delta)}_{\nu}}
  {1+(\rho/\rho_0)^2} + \lambda^{(\delta)}_{\nu}
  \label{e70}
  \;,\\
  &&
  \Big(\frac{\rho_0}{b}\Big)^2 = - \frac{2 a_B N(N-1)}
  {\lambda_{\nu}^{(f)} - \lambda^{(\delta)}_{\nu}} = \frac{2}{N-2}
  \label{e75}
  \;,
\end{eqnarray}
where the last equality only holds for $\nu=0$. Thus, $\rho_0$ is
independent of $a_B$ and inversely proportional to $N$.  In these
expressions we must for somewhat stronger attraction replace
$\lambda^{(\delta)}_{\nu}$ by the solution obtained numerically. The
curve for $a_B = -0.0559$ in fig.~\ref{fig2} suddenly decreases
dramatically when $\rho/b \approx 3000$. This reflects the pronounced
effect of the attractive pocket for large $\alpha$ appearing in
fig.~\ref{fig2} for large $\rho$.

The radial potential in eq.~(\ref{e35}) has the same simple structure
as the mean-field energy in eq.~(\ref{e55}). Analogously stable
solutions only exist for constant $\lambda$ when $\lambda >
-(N-3/2)(N-5/2)$. Using $\lambda^{(\delta)}_{\nu=0}$ this implies
collapse when $a_B < -(N-3/2)(N-5/2) /[N(N-1)(N-2)] \approx -1/(N+1)$.
This condition is again very similar to the mean-field condition in
eq.~(\ref{e77}), but now less surprising since the hyperradial and the
mean-field radial equations both describe the same overall size
dependence.

In fact a constant $\tilde \phi$ in eq.~(\ref{e30}) corresponding to
$\nu = 0$ in eq.~(\ref{e45}) leaves the total wave function as a
function of only the hyperradius $\rho$. For Gaussian single-particle
wave functions this is precisely identical to a Gaussian $\rho$
dependence of $f(\rho)$. For all other than Gaussian radial shapes of
single-particle and hyperradial wave functions the connection cannot
be made explicite.  In the present formulation correlations, and
thereby clear-cut deviations from the mean-field approximation, are
included through the angular dependence of $\Phi$. These effects
beyond the mean-field are perhaps most striking by the ability to
describe simultaneously both states resembling a condensate and
two-body bound states within the $N$-body system. This only requires
two adiabatic potentials.

For $\rho$ values confined by the external field, we can insert the
constant $\lambda^{(\delta)}_{\nu}$ given by eq.~(\ref{e70}) in
eq.~(\ref{e35}) which then is the radial harmonic-oscillator equation
for an effective angular-momentum quantum number $l^*$ defined by
$l^*(l^*+1) \equiv \lambda + (N-3/2)(N-5/2)$ or equivalently $l^* =
-0.5 + [\lambda + (N-3/2)(N-5/2) +1/4]^{1/2}$.  When $l^*(l^*+1) >
-1/4$, the solutions are well defined and characterized by the
corresponding real values of $l^*>-1/2$.  The eigenfunctions and
eigenvalues are $(\rho/b_t)^{l^*} L^{(l^*-1/2)}_n[(\rho/b_t)^2]
\exp[-0.5(\rho/b_t)^2]$ and $\hbar \omega (2n+l^*+3/2)$, where $L_n$
are generalized Laguerre polynomials and $n$ is a non-negative
integer.  The ground-state energy is then $\hbar \omega (l^*+3/2)$ and
the level spacing is precisely $2 \hbar \omega$ as for
harmonic-oscillator excitations maintaining parity.

\paragraph*{Renormalization of the interaction strength.}

In mean-field $3D$ zero-range calculations the employed interaction is
renormalized as $4\pi a_s \delta(\vec r) \hbar^2/m $, where $a_s$ is
the true $s$-wave scattering length derived from the low-energy
two-body scattering properties.  A short-range Gaussian representation
of the delta function with a strength adjusted to reproduce the same
scattering length $a_s$ leads to totally different mean-field results.
In order to get the desired results in the mean-field approximation
with a Gaussian interaction, it is necessary to renormalize the
Gaussian strength such that the Born approximation to the scattering
length is reproduced instead of the true scattering length
\cite{sor03b}.  Thus the interaction in $3D$ mean-field calculations
must reproduce the correct scattering length by using the Born
approximation.

This renormalization procedure guarantees three-dimensional mean-field
results in agreement with the use of the above renormalization of the
zero-range interaction.  In the hyperspherical adiabatic method the
large-distance asymptotic behavior of the angular eigenvalue should
approach the expectation value for the mean-field zero-range
interaction using the free wave function, i.e.~the proper
renormalization is seen to be equivalent to $\lambda_{\nu} \rightarrow
\lambda^{(\delta)}_{\nu}$ for large $\rho$.  Thus, to the extent the
large-$\rho$ behavior of $\lambda_{\nu}$ is well described by
$\lambda^{(\delta)}_{\nu}$, we conjecture that this renormalization is
not needed in $2D$ calculations.

Let us now average a $3D$ Gaussian interaction over the
$z$-coordinate.  Then we find that the Born approximation to the
scattering length of the new $2D$-interaction is $a_B \propto
a^{(3)}_B /b_z$, where $b_z$ is the trap length in the $z$-direction
and $b \ll b_z$ \cite{pit03}.  Considering refs.~\cite{zyl02,pet00}
and neglecting the logarithmic terms in their results, the present
result is precisely the same for zero-range interactions when the true
scattering length is substituted by the Born approximation.  Thus, in
$3D$ mean-field calculations the Born approximation to the scattering
length of the interaction potential, e.g.~a Gaussian, is replaced by
the true scattering length, whereas no replacement is needed in two
dimensions.

The stability condition, $N a_B > - 1$, is equivalent to $N
a^{(3)}_B/b_z > - 1$, which coincides with the condition obtained by
variation of the Gaussian length parameters for a deformed trap for a
zero-range interaction where the $3D$ scattering length is substituted
by its Born approximation \cite{bay96}.  Again this indicates that
renormalization is not needed for $2D$-calculations.

When the $3D$ scattering length is small compared to $b_z$, the $2D$
and $3D$ coupling strengths for a zero-range interaction are related
by $g_{2D}\approx g_{3D} /(b_z \sqrt{2\pi})$.  When the axial
confinement is strong, i.e.~$b_z \ll a_s$, $g_{2D}$ becomes density
dependent \cite{pet00}.

Treating correlations with the hyperspherical adiabatic method seems
to require renormalization only in three dimensions where the
mean-field and correlated results only coincide for differently
renormalized interactions.  For $2D$-calculations the same interaction
seems to produce the same large-distance mean-field and correlated
results. Thus, no renormalization seems to be required in $2D$.

\paragraph*{Conclusions.}

We investigated symmetric $N$-boson systems in the rigorous
two-dimensional limit for attractive two-body interactions.  We derive
the stability condition where the product of particle number,
strength, mass, and square of the range has to be sufficiently small.
The condition is the same for both mean-field and correlated solutions
and independent of the external field.  Diatomic recombination into
two-body bound states is not likely, since the two-body bound state
would extend beyond the trap length for potentials sufficiently weak
to allow stable solutions.  Thus, these weakly attractive potentials,
inevitably binding two particles, are able to support stable
condensates.  These features are completely different from
three-dimensional properties.  These rigorous two-dimensional results
cannot be compared to other calculations where the effective
interactions are repulsive.


\end{document}

%% file: hanl1fig1.tex
\begingroup%
  \makeatletter%
  \newcommand{\GNUPLOTspecial}{%
    \@sanitize\catcode`\%=14\relax\special}%
  \setlength{\unitlength}{0.1bp}%
{\GNUPLOTspecial{!
/gnudict 256 dict def
gnudict begin
/Color false def
/Solid false def
/gnulinewidth 5.000 def
/userlinewidth gnulinewidth def
/vshift -33 def
/dl {10 mul} def
/hpt_ 31.5 def
/vpt_ 31.5 def
/hpt hpt_ def
/vpt vpt_ def
/M {moveto} bind def
/L {lineto} bind def
/R {rmoveto} bind def
/V {rlineto} bind def
/vpt2 vpt 2 mul def
/hpt2 hpt 2 mul def
/Lshow { currentpoint stroke M
  0 vshift R show } def
/Rshow { currentpoint stroke M
  dup stringwidth pop neg vshift R show } def
/Cshow { currentpoint stroke M
  dup stringwidth pop -2 div vshift R show } def
/UP { dup vpt_ mul /vpt exch def hpt_ mul /hpt exch def
  /hpt2 hpt 2 mul def /vpt2 vpt 2 mul def } def
/DL { Color {setrgbcolor Solid {pop []} if 0 setdash }
 {pop pop pop Solid {pop []} if 0 setdash} ifelse } def
/BL { stroke userlinewidth 2 mul setlinewidth } def
/AL { stroke userlinewidth 2 div setlinewidth } def
/UL { dup gnulinewidth mul /userlinewidth exch def
      10 mul /udl exch def } def
/PL { stroke userlinewidth setlinewidth } def
/LTb { BL [] 0 0 0 DL } def
/LTa { AL [1 udl mul 2 udl mul] 0 setdash 0 0 0 setrgbcolor } def
/LT0 { PL [] 1 0 0 DL } def
/LT1 { PL [4 dl 2 dl] 0 1 0 DL } def
/LT2 { PL [2 dl 3 dl] 0 0 1 DL } def
/LT3 { PL [1 dl 1.5 dl] 1 0 1 DL } def
/LT4 { PL [5 dl 2 dl 1 dl 2 dl] 0 1 1 DL } def
/LT5 { PL [4 dl 3 dl 1 dl 3 dl] 1 1 0 DL } def
/LT6 { PL [2 dl 2 dl 2 dl 4 dl] 0 0 0 DL } def
/LT7 { PL [2 dl 2 dl 2 dl 2 dl 2 dl 4 dl] 1 0.3 0 DL } def
/LT8 { PL [2 dl 2 dl 2 dl 2 dl 2 dl 2 dl 2 dl 4 dl] 0.5 0.5 0.5 DL } def
/Pnt { stroke [] 0 setdash
   gsave 1 setlinecap M 0 0 V stroke grestore } def
/Dia { stroke [] 0 setdash 2 copy vpt add M
  hpt neg vpt neg V hpt vpt neg V
  hpt vpt V hpt neg vpt V closepath stroke
  Pnt } def
/Pls { stroke [] 0 setdash vpt sub M 0 vpt2 V
  currentpoint stroke M
  hpt neg vpt neg R hpt2 0 V stroke
  } def
/Box { stroke [] 0 setdash 2 copy exch hpt sub exch vpt add M
  0 vpt2 neg V hpt2 0 V 0 vpt2 V
  hpt2 neg 0 V closepath stroke
  Pnt } def
/Crs { stroke [] 0 setdash exch hpt sub exch vpt add M
  hpt2 vpt2 neg V currentpoint stroke M
  hpt2 neg 0 R hpt2 vpt2 V stroke } def
/TriU { stroke [] 0 setdash 2 copy vpt 1.12 mul add M
  hpt neg vpt -1.62 mul V
  hpt 2 mul 0 V
  hpt neg vpt 1.62 mul V closepath stroke
  Pnt  } def
/Star { 2 copy Pls Crs } def
/BoxF { stroke [] 0 setdash exch hpt sub exch vpt add M
  0 vpt2 neg V  hpt2 0 V  0 vpt2 V
  hpt2 neg 0 V  closepath fill } def
/TriUF { stroke [] 0 setdash vpt 1.12 mul add M
  hpt neg vpt -1.62 mul V
  hpt 2 mul 0 V
  hpt neg vpt 1.62 mul V closepath fill } def
/TriD { stroke [] 0 setdash 2 copy vpt 1.12 mul sub M
  hpt neg vpt 1.62 mul V
  hpt 2 mul 0 V
  hpt neg vpt -1.62 mul V closepath stroke
  Pnt  } def
/TriDF { stroke [] 0 setdash vpt 1.12 mul sub M
  hpt neg vpt 1.62 mul V
  hpt 2 mul 0 V
  hpt neg vpt -1.62 mul V closepath fill} def
/DiaF { stroke [] 0 setdash vpt add M
  hpt neg vpt neg V hpt vpt neg V
  hpt vpt V hpt neg vpt V closepath fill } def
/Pent { stroke [] 0 setdash 2 copy gsave
  translate 0 hpt M 4 {72 rotate 0 hpt L} repeat
  closepath stroke grestore Pnt } def
/PentF { stroke [] 0 setdash gsave
  translate 0 hpt M 4 {72 rotate 0 hpt L} repeat
  closepath fill grestore } def
/Circle { stroke [] 0 setdash 2 copy
  hpt 0 360 arc stroke Pnt } def
/CircleF { stroke [] 0 setdash hpt 0 360 arc fill } def
/C0 { BL [] 0 setdash 2 copy moveto vpt 90 450  arc } bind def
/C1 { BL [] 0 setdash 2 copy        moveto
       2 copy  vpt 0 90 arc closepath fill
               vpt 0 360 arc closepath } bind def
/C2 { BL [] 0 setdash 2 copy moveto
       2 copy  vpt 90 180 arc closepath fill
               vpt 0 360 arc closepath } bind def
/C3 { BL [] 0 setdash 2 copy moveto
       2 copy  vpt 0 180 arc closepath fill
               vpt 0 360 arc closepath } bind def
/C4 { BL [] 0 setdash 2 copy moveto
       2 copy  vpt 180 270 arc closepath fill
               vpt 0 360 arc closepath } bind def
/C5 { BL [] 0 setdash 2 copy moveto
       2 copy  vpt 0 90 arc
       2 copy moveto
       2 copy  vpt 180 270 arc closepath fill
               vpt 0 360 arc } bind def
/C6 { BL [] 0 setdash 2 copy moveto
      2 copy  vpt 90 270 arc closepath fill
              vpt 0 360 arc closepath } bind def
/C7 { BL [] 0 setdash 2 copy moveto
      2 copy  vpt 0 270 arc closepath fill
              vpt 0 360 arc closepath } bind def
/C8 { BL [] 0 setdash 2 copy moveto
      2 copy vpt 270 360 arc closepath fill
              vpt 0 360 arc closepath } bind def
/C9 { BL [] 0 setdash 2 copy moveto
      2 copy  vpt 270 450 arc closepath fill
              vpt 0 360 arc closepath } bind def
/C10 { BL [] 0 setdash 2 copy 2 copy moveto vpt 270 360 arc closepath fill
       2 copy moveto
       2 copy vpt 90 180 arc closepath fill
               vpt 0 360 arc closepath } bind def
/C11 { BL [] 0 setdash 2 copy moveto
       2 copy  vpt 0 180 arc closepath fill
       2 copy moveto
       2 copy  vpt 270 360 arc closepath fill
               vpt 0 360 arc closepath } bind def
/C12 { BL [] 0 setdash 2 copy moveto
       2 copy  vpt 180 360 arc closepath fill
               vpt 0 360 arc closepath } bind def
/C13 { BL [] 0 setdash  2 copy moveto
       2 copy  vpt 0 90 arc closepath fill
       2 copy moveto
       2 copy  vpt 180 360 arc closepath fill
               vpt 0 360 arc closepath } bind def
/C14 { BL [] 0 setdash 2 copy moveto
       2 copy  vpt 90 360 arc closepath fill
               vpt 0 360 arc } bind def
/C15 { BL [] 0 setdash 2 copy vpt 0 360 arc closepath fill
               vpt 0 360 arc closepath } bind def
/Rec   { newpath 4 2 roll moveto 1 index 0 rlineto 0 exch rlineto
       neg 0 rlineto closepath } bind def
/Square { dup Rec } bind def
/Bsquare { vpt sub exch vpt sub exch vpt2 Square } bind def
/S0 { BL [] 0 setdash 2 copy moveto 0 vpt rlineto BL Bsquare } bind def
/S1 { BL [] 0 setdash 2 copy vpt Square fill Bsquare } bind def
/S2 { BL [] 0 setdash 2 copy exch vpt sub exch vpt Square fill Bsquare } bind def
/S3 { BL [] 0 setdash 2 copy exch vpt sub exch vpt2 vpt Rec fill Bsquare } bind def
/S4 { BL [] 0 setdash 2 copy exch vpt sub exch vpt sub vpt Square fill Bsquare } bind def
/S5 { BL [] 0 setdash 2 copy 2 copy vpt Square fill
       exch vpt sub exch vpt sub vpt Square fill Bsquare } bind def
/S6 { BL [] 0 setdash 2 copy exch vpt sub exch vpt sub vpt vpt2 Rec fill Bsquare } bind def
/S7 { BL [] 0 setdash 2 copy exch vpt sub exch vpt sub vpt vpt2 Rec fill
       2 copy vpt Square fill
       Bsquare } bind def
/S8 { BL [] 0 setdash 2 copy vpt sub vpt Square fill Bsquare } bind def
/S9 { BL [] 0 setdash 2 copy vpt sub vpt vpt2 Rec fill Bsquare } bind def
/S10 { BL [] 0 setdash 2 copy vpt sub vpt Square fill 2 copy exch vpt sub exch vpt Square fill
       Bsquare } bind def
/S11 { BL [] 0 setdash 2 copy vpt sub vpt Square fill 2 copy exch vpt sub exch vpt2 vpt Rec fill
       Bsquare } bind def
/S12 { BL [] 0 setdash 2 copy exch vpt sub exch vpt sub vpt2 vpt Rec fill Bsquare } bind def
/S13 { BL [] 0 setdash 2 copy exch vpt sub exch vpt sub vpt2 vpt Rec fill
       2 copy vpt Square fill Bsquare } bind def
/S14 { BL [] 0 setdash 2 copy exch vpt sub exch vpt sub vpt2 vpt Rec fill
       2 copy exch vpt sub exch vpt Square fill Bsquare } bind def
/S15 { BL [] 0 setdash 2 copy Bsquare fill Bsquare } bind def
/D0 { gsave translate 45 rotate 0 0 S0 stroke grestore } bind def
/D1 { gsave translate 45 rotate 0 0 S1 stroke grestore } bind def
/D2 { gsave translate 45 rotate 0 0 S2 stroke grestore } bind def
/D3 { gsave translate 45 rotate 0 0 S3 stroke grestore } bind def
/D4 { gsave translate 45 rotate 0 0 S4 stroke grestore } bind def
/D5 { gsave translate 45 rotate 0 0 S5 stroke grestore } bind def
/D6 { gsave translate 45 rotate 0 0 S6 stroke grestore } bind def
/D7 { gsave translate 45 rotate 0 0 S7 stroke grestore } bind def
/D8 { gsave translate 45 rotate 0 0 S8 stroke grestore } bind def
/D9 { gsave translate 45 rotate 0 0 S9 stroke grestore } bind def
/D10 { gsave translate 45 rotate 0 0 S10 stroke grestore } bind def
/D11 { gsave translate 45 rotate 0 0 S11 stroke grestore } bind def
/D12 { gsave translate 45 rotate 0 0 S12 stroke grestore } bind def
/D13 { gsave translate 45 rotate 0 0 S13 stroke grestore } bind def
/D14 { gsave translate 45 rotate 0 0 S14 stroke grestore } bind def
/D15 { gsave translate 45 rotate 0 0 S15 stroke grestore } bind def
/DiaE { stroke [] 0 setdash vpt add M
  hpt neg vpt neg V hpt vpt neg V
  hpt vpt V hpt neg vpt V closepath stroke } def
/BoxE { stroke [] 0 setdash exch hpt sub exch vpt add M
  0 vpt2 neg V hpt2 0 V 0 vpt2 V
  hpt2 neg 0 V closepath stroke } def
/TriUE { stroke [] 0 setdash vpt 1.12 mul add M
  hpt neg vpt -1.62 mul V
  hpt 2 mul 0 V
  hpt neg vpt 1.62 mul V closepath stroke } def
/TriDE { stroke [] 0 setdash vpt 1.12 mul sub M
  hpt neg vpt 1.62 mul V
  hpt 2 mul 0 V
  hpt neg vpt -1.62 mul V closepath stroke } def
/PentE { stroke [] 0 setdash gsave
  translate 0 hpt M 4 {72 rotate 0 hpt L} repeat
  closepath stroke grestore } def
/CircE { stroke [] 0 setdash 
  hpt 0 360 arc stroke } def
/Opaque { gsave closepath 1 setgray fill grestore 0 setgray closepath } def
/DiaW { stroke [] 0 setdash vpt add M
  hpt neg vpt neg V hpt vpt neg V
  hpt vpt V hpt neg vpt V Opaque stroke } def
/BoxW { stroke [] 0 setdash exch hpt sub exch vpt add M
  0 vpt2 neg V hpt2 0 V 0 vpt2 V
  hpt2 neg 0 V Opaque stroke } def
/TriUW { stroke [] 0 setdash vpt 1.12 mul add M
  hpt neg vpt -1.62 mul V
  hpt 2 mul 0 V
  hpt neg vpt 1.62 mul V Opaque stroke } def
/TriDW { stroke [] 0 setdash vpt 1.12 mul sub M
  hpt neg vpt 1.62 mul V
  hpt 2 mul 0 V
  hpt neg vpt -1.62 mul V Opaque stroke } def
/PentW { stroke [] 0 setdash gsave
  translate 0 hpt M 4 {72 rotate 0 hpt L} repeat
  Opaque stroke grestore } def
/CircW { stroke [] 0 setdash 
  hpt 0 360 arc Opaque stroke } def
/BoxFill { gsave Rec 1 setgray fill grestore } def
end
}}%
\begin{picture}(2448,2376)(0,0)%
{\GNUPLOTspecial{"
gnudict begin
gsave
0 0 translate
0.100 0.100 scale
0 setgray
newpath
1.000 UL
LTb
500 300 M
63 0 V
1735 0 R
-63 0 V
500 359 M
31 0 V
1767 0 R
-31 0 V
500 418 M
31 0 V
1767 0 R
-31 0 V
500 478 M
31 0 V
1767 0 R
-31 0 V
500 537 M
63 0 V
1735 0 R
-63 0 V
500 596 M
31 0 V
1767 0 R
-31 0 V
500 655 M
31 0 V
1767 0 R
-31 0 V
500 714 M
31 0 V
1767 0 R
-31 0 V
500 774 M
63 0 V
1735 0 R
-63 0 V
500 833 M
31 0 V
1767 0 R
-31 0 V
500 892 M
31 0 V
1767 0 R
-31 0 V
500 951 M
31 0 V
1767 0 R
-31 0 V
500 1010 M
63 0 V
1735 0 R
-63 0 V
500 1070 M
31 0 V
1767 0 R
-31 0 V
500 1129 M
31 0 V
1767 0 R
-31 0 V
500 1188 M
31 0 V
1767 0 R
-31 0 V
500 1247 M
63 0 V
1735 0 R
-63 0 V
500 1306 M
31 0 V
1767 0 R
-31 0 V
500 1366 M
31 0 V
1767 0 R
-31 0 V
500 1425 M
31 0 V
1767 0 R
-31 0 V
500 1484 M
63 0 V
1735 0 R
-63 0 V
500 1543 M
31 0 V
1767 0 R
-31 0 V
500 1602 M
31 0 V
1767 0 R
-31 0 V
500 1662 M
31 0 V
1767 0 R
-31 0 V
500 1721 M
63 0 V
1735 0 R
-63 0 V
500 1780 M
31 0 V
1767 0 R
-31 0 V
500 1839 M
31 0 V
1767 0 R
-31 0 V
500 1898 M
31 0 V
1767 0 R
-31 0 V
500 1958 M
63 0 V
1735 0 R
-63 0 V
500 2017 M
31 0 V
1767 0 R
-31 0 V
500 2076 M
31 0 V
1767 0 R
-31 0 V
500 300 M
0 63 V
0 1713 R
0 -63 V
614 300 M
0 31 V
0 1745 R
0 -31 V
729 300 M
0 63 V
0 1713 R
0 -63 V
843 300 M
0 31 V
0 1745 R
0 -31 V
958 300 M
0 63 V
0 1713 R
0 -63 V
1072 300 M
0 31 V
0 1745 R
0 -31 V
1187 300 M
0 63 V
0 1713 R
0 -63 V
1301 300 M
0 31 V
0 1745 R
0 -31 V
1416 300 M
0 63 V
0 1713 R
0 -63 V
1530 300 M
0 31 V
0 1745 R
0 -31 V
1644 300 M
0 63 V
0 1713 R
0 -63 V
1759 300 M
0 31 V
0 1745 R
0 -31 V
1873 300 M
0 63 V
0 1713 R
0 -63 V
1988 300 M
0 31 V
0 1745 R
0 -31 V
2102 300 M
0 63 V
0 1713 R
0 -63 V
2217 300 M
0 31 V
0 1745 R
0 -31 V
1.000 UL
LTb
500 300 M
1798 0 V
0 1776 V
-1798 0 V
500 300 L
1.000 UL
LT0
500 1484 M
500 300 L
19 0 R
1 201 V
6 365 V
7 225 V
10 137 V
11 84 V
16 52 V
19 32 V
26 20 V
32 14 V
42 11 V
48 9 V
48 9 V
48 10 V
48 11 V
48 13 V
48 14 V
48 17 V
48 19 V
48 22 V
48 26 V
48 30 V
48 34 V
48 40 V
48 46 V
48 55 V
48 65 V
48 78 V
48 93 V
19 44 V
1.000 UL
LT0
500 1361 M
1 0 V
1 0 V
1 0 V
1 0 V
1 0 V
1 0 V
2 0 V
2 0 V
2 0 V
4 0 V
4 0 V
6 0 V
7 0 V
10 0 V
11 0 V
16 -1 V
19 0 V
26 0 V
32 -1 V
42 -2 V
48 -2 V
48 -2 V
48 -3 V
48 -4 V
48 -4 V
48 -5 V
48 -6 V
48 -7 V
48 -8 V
48 -9 V
48 -11 V
48 -12 V
48 -14 V
48 -17 V
48 -20 V
48 -23 V
48 -28 V
48 -34 V
48 -41 V
48 -51 V
48 -63 V
48 -81 V
48 -105 V
48 -140 V
48 -193 V
31 -174 V
4.000 UL
LT0
600 613 M
263 0 V
500 1361 M
500 300 L
20 0 R
0 78 V
6 365 V
7 224 V
10 138 V
11 84 V
16 51 V
19 32 V
26 20 V
32 13 V
42 9 V
48 8 V
48 6 V
48 7 V
48 7 V
48 9 V
48 9 V
48 11 V
48 12 V
48 15 V
48 16 V
48 19 V
48 22 V
48 25 V
48 30 V
48 35 V
48 42 V
48 49 V
48 60 V
48 73 V
48 90 V
48 113 V
35 104 V
4.000 UL
LT1
600 513 M
263 0 V
500 1140 M
0 -840 V
23 0 R
3 222 V
7 224 V
10 137 V
11 84 V
16 51 V
19 31 V
26 19 V
32 12 V
42 7 V
48 4 V
48 2 V
48 1 V
48 1 V
48 0 V
48 1 V
48 0 V
48 0 V
48 0 V
48 0 V
48 0 V
48 0 V
48 0 V
48 -1 V
48 0 V
48 -1 V
48 0 V
48 -1 V
48 0 V
48 -1 V
48 -2 V
48 -1 V
48 -2 V
48 -3 V
48 -4 V
48 -5 V
49 -8 V
48 -12 V
48 -21 V
48 -41 V
16 -21 V
8 -14 V
8 -16 V
8 -18 V
8 -21 V
8 -25 V
8 -30 V
8 -36 V
8 -42 V
8 -49 V
8 -55 V
8 -57 V
8 -38 V
8 43 V
8 343 V
6 1119 V
4.000 UL
LT1
500 1140 M
0 -840 V
22 0 R
2 130 V
4 179 V
5 133 V
6 98 V
7 73 V
8 52 V
10 38 V
12 28 V
14 19 V
17 14 V
20 10 V
25 7 V
28 5 V
35 4 V
42 2 V
46 2 V
46 1 V
46 0 V
46 1 V
46 0 V
46 0 V
46 0 V
46 0 V
46 0 V
46 0 V
46 0 V
46 0 V
46 0 V
46 0 V
46 0 V
46 0 V
46 -1 V
46 0 V
46 0 V
46 0 V
45 0 V
47 0 V
46 0 V
46 2 V
46 3 V
45 8 V
47 17 V
46 42 V
46 115 V
46 381 V
16 301 V
3 112 V
4.000 UL
LT3
600 413 M
263 0 V
500 1079 M
0 -779 V
23 0 R
3 161 V
7 224 V
10 137 V
11 84 V
16 51 V
19 31 V
26 19 V
32 11 V
42 6 V
48 3 V
48 1 V
48 0 V
48 -1 V
48 -2 V
48 -2 V
48 -3 V
48 -3 V
48 -4 V
48 -5 V
48 -5 V
48 -6 V
48 -8 V
48 -8 V
48 -10 V
48 -12 V
48 -15 V
48 -17 V
48 -21 V
48 -26 V
48 -33 V
48 -41 V
48 -54 V
48 -73 V
48 -99 V
48 -141 V
32 -139 V
stroke
grestore
end
showpage
}}%
\put(913,413){\makebox(0,0)[l]{$a_B=-0.0658$}}%
\put(913,513){\makebox(0,0)[l]{$a_B=-0.0559$}}%
\put(913,613){\makebox(0,0)[l]{$a_B=-0.02$}}%
\put(637,1839){\makebox(0,0)[l]{$\rho/b=10^3$}}%
\put(637,1958){\makebox(0,0)[l]{$N=20$}}%
\put(1759,1543){\makebox(0,0)[l]{\scriptsize$\rho/b=10^2$}}%
\put(1438,1271){\makebox(0,0)[l]{$v_1$}}%
\put(1187,1721){\makebox(0,0)[l]{$v_c$}}%
\put(1399,2226){\makebox(0,0){ }}%
\put(1399,50){\makebox(0,0){$\alpha $}}%
\put(100,1188){%
\special{ps: gsave currentpoint currentpoint translate
270 rotate neg exch neg exch translate}%
\makebox(0,0)[b]{\shortstack{$V$}}%
\special{ps: currentpoint grestore moveto}%
}%
\put(2102,200){\makebox(0,0){1.4}}%
\put(1873,200){\makebox(0,0){1.2}}%
\put(1644,200){\makebox(0,0){1}}%
\put(1416,200){\makebox(0,0){0.8}}%
\put(1187,200){\makebox(0,0){0.6}}%
\put(958,200){\makebox(0,0){0.4}}%
\put(729,200){\makebox(0,0){0.2}}%
\put(500,200){\makebox(0,0){0}}%
\put(450,1958){\makebox(0,0)[r]{400}}%
\put(450,1721){\makebox(0,0)[r]{200}}%
\put(450,1484){\makebox(0,0)[r]{0}}%
\put(450,1247){\makebox(0,0)[r]{-200}}%
\put(450,1010){\makebox(0,0)[r]{-400}}%
\put(450,774){\makebox(0,0)[r]{-600}}%
\put(450,537){\makebox(0,0)[r]{-800}}%
\put(450,300){\makebox(0,0)[r]{-1000}}%
\end{picture}%
\endgroup
 

%% file: hanl1fig2.tex
\begingroup%
  \makeatletter%
  \newcommand{\GNUPLOTspecial}{%
    \@sanitize\catcode`\%=14\relax\special}%
  \setlength{\unitlength}{0.1bp}%
{\GNUPLOTspecial{!
/gnudict 256 dict def
gnudict begin
/Color false def
/Solid false def
/gnulinewidth 5.000 def
/userlinewidth gnulinewidth def
/vshift -33 def
/dl {10 mul} def
/hpt_ 31.5 def
/vpt_ 31.5 def
/hpt hpt_ def
/vpt vpt_ def
/M {moveto} bind def
/L {lineto} bind def
/R {rmoveto} bind def
/V {rlineto} bind def
/vpt2 vpt 2 mul def
/hpt2 hpt 2 mul def
/Lshow { currentpoint stroke M
  0 vshift R show } def
/Rshow { currentpoint stroke M
  dup stringwidth pop neg vshift R show } def
/Cshow { currentpoint stroke M
  dup stringwidth pop -2 div vshift R show } def
/UP { dup vpt_ mul /vpt exch def hpt_ mul /hpt exch def
  /hpt2 hpt 2 mul def /vpt2 vpt 2 mul def } def
/DL { Color {setrgbcolor Solid {pop []} if 0 setdash }
 {pop pop pop Solid {pop []} if 0 setdash} ifelse } def
/BL { stroke userlinewidth 2 mul setlinewidth } def
/AL { stroke userlinewidth 2 div setlinewidth } def
/UL { dup gnulinewidth mul /userlinewidth exch def
      10 mul /udl exch def } def
/PL { stroke userlinewidth setlinewidth } def
/LTb { BL [] 0 0 0 DL } def
/LTa { AL [1 udl mul 2 udl mul] 0 setdash 0 0 0 setrgbcolor } def
/LT0 { PL [] 1 0 0 DL } def
/LT1 { PL [4 dl 2 dl] 0 1 0 DL } def
/LT2 { PL [2 dl 3 dl] 0 0 1 DL } def
/LT3 { PL [1 dl 1.5 dl] 1 0 1 DL } def
/LT4 { PL [5 dl 2 dl 1 dl 2 dl] 0 1 1 DL } def
/LT5 { PL [4 dl 3 dl 1 dl 3 dl] 1 1 0 DL } def
/LT6 { PL [2 dl 2 dl 2 dl 4 dl] 0 0 0 DL } def
/LT7 { PL [2 dl 2 dl 2 dl 2 dl 2 dl 4 dl] 1 0.3 0 DL } def
/LT8 { PL [2 dl 2 dl 2 dl 2 dl 2 dl 2 dl 2 dl 4 dl] 0.5 0.5 0.5 DL } def
/Pnt { stroke [] 0 setdash
   gsave 1 setlinecap M 0 0 V stroke grestore } def
/Dia { stroke [] 0 setdash 2 copy vpt add M
  hpt neg vpt neg V hpt vpt neg V
  hpt vpt V hpt neg vpt V closepath stroke
  Pnt } def
/Pls { stroke [] 0 setdash vpt sub M 0 vpt2 V
  currentpoint stroke M
  hpt neg vpt neg R hpt2 0 V stroke
  } def
/Box { stroke [] 0 setdash 2 copy exch hpt sub exch vpt add M
  0 vpt2 neg V hpt2 0 V 0 vpt2 V
  hpt2 neg 0 V closepath stroke
  Pnt } def
/Crs { stroke [] 0 setdash exch hpt sub exch vpt add M
  hpt2 vpt2 neg V currentpoint stroke M
  hpt2 neg 0 R hpt2 vpt2 V stroke } def
/TriU { stroke [] 0 setdash 2 copy vpt 1.12 mul add M
  hpt neg vpt -1.62 mul V
  hpt 2 mul 0 V
  hpt neg vpt 1.62 mul V closepath stroke
  Pnt  } def
/Star { 2 copy Pls Crs } def
/BoxF { stroke [] 0 setdash exch hpt sub exch vpt add M
  0 vpt2 neg V  hpt2 0 V  0 vpt2 V
  hpt2 neg 0 V  closepath fill } def
/TriUF { stroke [] 0 setdash vpt 1.12 mul add M
  hpt neg vpt -1.62 mul V
  hpt 2 mul 0 V
  hpt neg vpt 1.62 mul V closepath fill } def
/TriD { stroke [] 0 setdash 2 copy vpt 1.12 mul sub M
  hpt neg vpt 1.62 mul V
  hpt 2 mul 0 V
  hpt neg vpt -1.62 mul V closepath stroke
  Pnt  } def
/TriDF { stroke [] 0 setdash vpt 1.12 mul sub M
  hpt neg vpt 1.62 mul V
  hpt 2 mul 0 V
  hpt neg vpt -1.62 mul V closepath fill} def
/DiaF { stroke [] 0 setdash vpt add M
  hpt neg vpt neg V hpt vpt neg V
  hpt vpt V hpt neg vpt V closepath fill } def
/Pent { stroke [] 0 setdash 2 copy gsave
  translate 0 hpt M 4 {72 rotate 0 hpt L} repeat
  closepath stroke grestore Pnt } def
/PentF { stroke [] 0 setdash gsave
  translate 0 hpt M 4 {72 rotate 0 hpt L} repeat
  closepath fill grestore } def
/Circle { stroke [] 0 setdash 2 copy
  hpt 0 360 arc stroke Pnt } def
/CircleF { stroke [] 0 setdash hpt 0 360 arc fill } def
/C0 { BL [] 0 setdash 2 copy moveto vpt 90 450  arc } bind def
/C1 { BL [] 0 setdash 2 copy        moveto
       2 copy  vpt 0 90 arc closepath fill
               vpt 0 360 arc closepath } bind def
/C2 { BL [] 0 setdash 2 copy moveto
       2 copy  vpt 90 180 arc closepath fill
               vpt 0 360 arc closepath } bind def
/C3 { BL [] 0 setdash 2 copy moveto
       2 copy  vpt 0 180 arc closepath fill
               vpt 0 360 arc closepath } bind def
/C4 { BL [] 0 setdash 2 copy moveto
       2 copy  vpt 180 270 arc closepath fill
               vpt 0 360 arc closepath } bind def
/C5 { BL [] 0 setdash 2 copy moveto
       2 copy  vpt 0 90 arc
       2 copy moveto
       2 copy  vpt 180 270 arc closepath fill
               vpt 0 360 arc } bind def
/C6 { BL [] 0 setdash 2 copy moveto
      2 copy  vpt 90 270 arc closepath fill
              vpt 0 360 arc closepath } bind def
/C7 { BL [] 0 setdash 2 copy moveto
      2 copy  vpt 0 270 arc closepath fill
              vpt 0 360 arc closepath } bind def
/C8 { BL [] 0 setdash 2 copy moveto
      2 copy vpt 270 360 arc closepath fill
              vpt 0 360 arc closepath } bind def
/C9 { BL [] 0 setdash 2 copy moveto
      2 copy  vpt 270 450 arc closepath fill
              vpt 0 360 arc closepath } bind def
/C10 { BL [] 0 setdash 2 copy 2 copy moveto vpt 270 360 arc closepath fill
       2 copy moveto
       2 copy vpt 90 180 arc closepath fill
               vpt 0 360 arc closepath } bind def
/C11 { BL [] 0 setdash 2 copy moveto
       2 copy  vpt 0 180 arc closepath fill
       2 copy moveto
       2 copy  vpt 270 360 arc closepath fill
               vpt 0 360 arc closepath } bind def
/C12 { BL [] 0 setdash 2 copy moveto
       2 copy  vpt 180 360 arc closepath fill
               vpt 0 360 arc closepath } bind def
/C13 { BL [] 0 setdash  2 copy moveto
       2 copy  vpt 0 90 arc closepath fill
       2 copy moveto
       2 copy  vpt 180 360 arc closepath fill
               vpt 0 360 arc closepath } bind def
/C14 { BL [] 0 setdash 2 copy moveto
       2 copy  vpt 90 360 arc closepath fill
               vpt 0 360 arc } bind def
/C15 { BL [] 0 setdash 2 copy vpt 0 360 arc closepath fill
               vpt 0 360 arc closepath } bind def
/Rec   { newpath 4 2 roll moveto 1 index 0 rlineto 0 exch rlineto
       neg 0 rlineto closepath } bind def
/Square { dup Rec } bind def
/Bsquare { vpt sub exch vpt sub exch vpt2 Square } bind def
/S0 { BL [] 0 setdash 2 copy moveto 0 vpt rlineto BL Bsquare } bind def
/S1 { BL [] 0 setdash 2 copy vpt Square fill Bsquare } bind def
/S2 { BL [] 0 setdash 2 copy exch vpt sub exch vpt Square fill Bsquare } bind def
/S3 { BL [] 0 setdash 2 copy exch vpt sub exch vpt2 vpt Rec fill Bsquare } bind def
/S4 { BL [] 0 setdash 2 copy exch vpt sub exch vpt sub vpt Square fill Bsquare } bind def
/S5 { BL [] 0 setdash 2 copy 2 copy vpt Square fill
       exch vpt sub exch vpt sub vpt Square fill Bsquare } bind def
/S6 { BL [] 0 setdash 2 copy exch vpt sub exch vpt sub vpt vpt2 Rec fill Bsquare } bind def
/S7 { BL [] 0 setdash 2 copy exch vpt sub exch vpt sub vpt vpt2 Rec fill
       2 copy vpt Square fill
       Bsquare } bind def
/S8 { BL [] 0 setdash 2 copy vpt sub vpt Square fill Bsquare } bind def
/S9 { BL [] 0 setdash 2 copy vpt sub vpt vpt2 Rec fill Bsquare } bind def
/S10 { BL [] 0 setdash 2 copy vpt sub vpt Square fill 2 copy exch vpt sub exch vpt Square fill
       Bsquare } bind def
/S11 { BL [] 0 setdash 2 copy vpt sub vpt Square fill 2 copy exch vpt sub exch vpt2 vpt Rec fill
       Bsquare } bind def
/S12 { BL [] 0 setdash 2 copy exch vpt sub exch vpt sub vpt2 vpt Rec fill Bsquare } bind def
/S13 { BL [] 0 setdash 2 copy exch vpt sub exch vpt sub vpt2 vpt Rec fill
       2 copy vpt Square fill Bsquare } bind def
/S14 { BL [] 0 setdash 2 copy exch vpt sub exch vpt sub vpt2 vpt Rec fill
       2 copy exch vpt sub exch vpt Square fill Bsquare } bind def
/S15 { BL [] 0 setdash 2 copy Bsquare fill Bsquare } bind def
/D0 { gsave translate 45 rotate 0 0 S0 stroke grestore } bind def
/D1 { gsave translate 45 rotate 0 0 S1 stroke grestore } bind def
/D2 { gsave translate 45 rotate 0 0 S2 stroke grestore } bind def
/D3 { gsave translate 45 rotate 0 0 S3 stroke grestore } bind def
/D4 { gsave translate 45 rotate 0 0 S4 stroke grestore } bind def
/D5 { gsave translate 45 rotate 0 0 S5 stroke grestore } bind def
/D6 { gsave translate 45 rotate 0 0 S6 stroke grestore } bind def
/D7 { gsave translate 45 rotate 0 0 S7 stroke grestore } bind def
/D8 { gsave translate 45 rotate 0 0 S8 stroke grestore } bind def
/D9 { gsave translate 45 rotate 0 0 S9 stroke grestore } bind def
/D10 { gsave translate 45 rotate 0 0 S10 stroke grestore } bind def
/D11 { gsave translate 45 rotate 0 0 S11 stroke grestore } bind def
/D12 { gsave translate 45 rotate 0 0 S12 stroke grestore } bind def
/D13 { gsave translate 45 rotate 0 0 S13 stroke grestore } bind def
/D14 { gsave translate 45 rotate 0 0 S14 stroke grestore } bind def
/D15 { gsave translate 45 rotate 0 0 S15 stroke grestore } bind def
/DiaE { stroke [] 0 setdash vpt add M
  hpt neg vpt neg V hpt vpt neg V
  hpt vpt V hpt neg vpt V closepath stroke } def
/BoxE { stroke [] 0 setdash exch hpt sub exch vpt add M
  0 vpt2 neg V hpt2 0 V 0 vpt2 V
  hpt2 neg 0 V closepath stroke } def
/TriUE { stroke [] 0 setdash vpt 1.12 mul add M
  hpt neg vpt -1.62 mul V
  hpt 2 mul 0 V
  hpt neg vpt 1.62 mul V closepath stroke } def
/TriDE { stroke [] 0 setdash vpt 1.12 mul sub M
  hpt neg vpt 1.62 mul V
  hpt 2 mul 0 V
  hpt neg vpt -1.62 mul V closepath stroke } def
/PentE { stroke [] 0 setdash gsave
  translate 0 hpt M 4 {72 rotate 0 hpt L} repeat
  closepath stroke grestore } def
/CircE { stroke [] 0 setdash 
  hpt 0 360 arc stroke } def
/Opaque { gsave closepath 1 setgray fill grestore 0 setgray closepath } def
/DiaW { stroke [] 0 setdash vpt add M
  hpt neg vpt neg V hpt vpt neg V
  hpt vpt V hpt neg vpt V Opaque stroke } def
/BoxW { stroke [] 0 setdash exch hpt sub exch vpt add M
  0 vpt2 neg V hpt2 0 V 0 vpt2 V
  hpt2 neg 0 V Opaque stroke } def
/TriUW { stroke [] 0 setdash vpt 1.12 mul add M
  hpt neg vpt -1.62 mul V
  hpt 2 mul 0 V
  hpt neg vpt 1.62 mul V Opaque stroke } def
/TriDW { stroke [] 0 setdash vpt 1.12 mul sub M
  hpt neg vpt 1.62 mul V
  hpt 2 mul 0 V
  hpt neg vpt -1.62 mul V Opaque stroke } def
/PentW { stroke [] 0 setdash gsave
  translate 0 hpt M 4 {72 rotate 0 hpt L} repeat
  Opaque stroke grestore } def
/CircW { stroke [] 0 setdash 
  hpt 0 360 arc Opaque stroke } def
/BoxFill { gsave Rec 1 setgray fill grestore } def
end
}}%
\begin{picture}(2448,2376)(0,0)%
{\GNUPLOTspecial{"
gnudict begin
gsave
0 0 translate
0.100 0.100 scale
0 setgray
newpath
1.000 UL
LTb
450 300 M
31 0 V
1817 0 R
-31 0 V
450 352 M
31 0 V
1817 0 R
-31 0 V
450 404 M
63 0 V
1785 0 R
-63 0 V
450 457 M
31 0 V
1817 0 R
-31 0 V
450 509 M
31 0 V
1817 0 R
-31 0 V
450 561 M
31 0 V
1817 0 R
-31 0 V
450 613 M
63 0 V
1785 0 R
-63 0 V
450 666 M
31 0 V
1817 0 R
-31 0 V
450 718 M
31 0 V
1817 0 R
-31 0 V
450 770 M
31 0 V
1817 0 R
-31 0 V
450 822 M
63 0 V
1785 0 R
-63 0 V
450 875 M
31 0 V
1817 0 R
-31 0 V
450 927 M
31 0 V
1817 0 R
-31 0 V
450 979 M
31 0 V
1817 0 R
-31 0 V
450 1031 M
63 0 V
1785 0 R
-63 0 V
450 1084 M
31 0 V
1817 0 R
-31 0 V
450 1136 M
31 0 V
1817 0 R
-31 0 V
450 1188 M
31 0 V
1817 0 R
-31 0 V
450 1240 M
63 0 V
1785 0 R
-63 0 V
450 1292 M
31 0 V
1817 0 R
-31 0 V
450 1345 M
31 0 V
1817 0 R
-31 0 V
450 1397 M
31 0 V
1817 0 R
-31 0 V
450 1449 M
63 0 V
1785 0 R
-63 0 V
450 1501 M
31 0 V
1817 0 R
-31 0 V
450 1554 M
31 0 V
1817 0 R
-31 0 V
450 1606 M
31 0 V
1817 0 R
-31 0 V
450 1658 M
63 0 V
1785 0 R
-63 0 V
450 1710 M
31 0 V
1817 0 R
-31 0 V
450 1763 M
31 0 V
1817 0 R
-31 0 V
450 1815 M
31 0 V
1817 0 R
-31 0 V
450 1867 M
63 0 V
1785 0 R
-63 0 V
450 1919 M
31 0 V
1817 0 R
-31 0 V
450 1972 M
31 0 V
1817 0 R
-31 0 V
450 2024 M
31 0 V
1817 0 R
-31 0 V
450 2076 M
63 0 V
1785 0 R
-63 0 V
450 300 M
0 63 V
0 1713 R
0 -63 V
543 300 M
0 31 V
0 1745 R
0 -31 V
665 300 M
0 31 V
0 1745 R
0 -31 V
728 300 M
0 31 V
0 1745 R
0 -31 V
758 300 M
0 63 V
0 1713 R
0 -63 V
851 300 M
0 31 V
0 1745 R
0 -31 V
973 300 M
0 31 V
0 1745 R
0 -31 V
1036 300 M
0 31 V
0 1745 R
0 -31 V
1066 300 M
0 63 V
0 1713 R
0 -63 V
1159 300 M
0 31 V
0 1745 R
0 -31 V
1281 300 M
0 31 V
0 1745 R
0 -31 V
1344 300 M
0 31 V
0 1745 R
0 -31 V
1374 300 M
0 63 V
0 1713 R
0 -63 V
1467 300 M
0 31 V
0 1745 R
0 -31 V
1589 300 M
0 31 V
0 1745 R
0 -31 V
1652 300 M
0 31 V
0 1745 R
0 -31 V
1682 300 M
0 63 V
0 1713 R
0 -63 V
1775 300 M
0 31 V
0 1745 R
0 -31 V
1897 300 M
0 31 V
0 1745 R
0 -31 V
1960 300 M
0 31 V
0 1745 R
0 -31 V
1990 300 M
0 63 V
0 1713 R
0 -63 V
2083 300 M
0 31 V
0 1745 R
0 -31 V
2205 300 M
0 31 V
0 1745 R
0 -31 V
2268 300 M
0 31 V
0 1745 R
0 -31 V
2298 300 M
0 63 V
0 1713 R
0 -63 V
1.000 UL
LTb
450 300 M
1848 0 V
0 1776 V
-1848 0 V
450 300 L
4.000 UL
LT0
450 1668 M
32 0 V
33 0 V
32 -1 V
33 -1 V
32 -2 V
33 -2 V
32 -4 V
32 -7 V
33 -10 V
32 -15 V
33 -21 V
32 -30 V
32 -36 V
33 -42 V
32 -41 V
33 -36 V
32 -30 V
33 -21 V
32 -15 V
32 -10 V
31 -6 V
32 -4 V
31 -3 V
32 -2 V
32 -1 V
31 -1 V
32 0 V
31 -1 V
32 0 V
31 0 V
32 0 V
32 -1 V
31 0 V
32 0 V
31 0 V
32 0 V
32 0 V
31 0 V
32 0 V
31 0 V
32 0 V
32 -1 V
31 0 V
32 0 V
31 0 V
32 0 V
32 0 V
31 0 V
32 0 V
31 0 V
32 0 V
31 0 V
32 0 V
32 0 V
31 0 V
32 0 V
31 -1 V
32 0 V
1.000 UL
LT0
450 1829 M
32 0 V
33 0 V
32 -1 V
33 -1 V
32 -1 V
33 -3 V
32 -3 V
32 -6 V
33 -9 V
32 -13 V
33 -20 V
32 -27 V
32 -35 V
33 -39 V
32 -40 V
33 -37 V
32 -30 V
33 -22 V
32 -16 V
32 -11 V
31 -7 V
32 -4 V
31 -3 V
32 -2 V
32 -1 V
31 -1 V
32 -1 V
31 0 V
32 0 V
31 -1 V
32 0 V
32 0 V
31 0 V
32 0 V
31 0 V
32 0 V
32 0 V
31 -1 V
32 0 V
31 0 V
32 0 V
32 0 V
31 0 V
32 0 V
31 0 V
32 0 V
32 0 V
31 0 V
32 0 V
31 -1 V
32 0 V
31 0 V
32 0 V
32 0 V
31 0 V
32 0 V
31 0 V
32 0 V
1.000 UL
LT0
450 2006 M
32 0 V
33 0 V
32 -1 V
33 0 V
32 -2 V
33 -2 V
32 -4 V
32 -5 V
33 -9 V
32 -13 V
33 -18 V
32 -27 V
32 -34 V
33 -40 V
32 -41 V
33 -39 V
32 -32 V
33 -25 V
32 -18 V
32 -13 V
31 -8 V
32 -5 V
31 -3 V
32 -3 V
32 -1 V
31 -1 V
32 -1 V
31 0 V
32 -1 V
31 0 V
32 0 V
32 0 V
31 0 V
32 -1 V
31 0 V
32 0 V
32 0 V
31 0 V
32 0 V
31 0 V
32 0 V
32 0 V
31 -1 V
32 0 V
31 0 V
32 0 V
32 0 V
31 0 V
32 0 V
31 0 V
32 0 V
31 0 V
32 0 V
32 0 V
31 0 V
32 0 V
31 0 V
32 -1 V
1.000 UL
LT0
879 2076 M
25 -31 V
32 -44 V
33 -42 V
32 -36 V
33 -28 V
32 -21 V
32 -15 V
31 -9 V
32 -7 V
31 -4 V
32 -3 V
32 -2 V
31 -1 V
32 -1 V
31 0 V
32 -1 V
31 0 V
32 0 V
32 0 V
31 -1 V
32 0 V
31 0 V
32 0 V
32 0 V
31 0 V
32 0 V
31 0 V
32 -1 V
32 0 V
31 0 V
32 0 V
31 0 V
32 0 V
32 0 V
31 0 V
32 0 V
31 0 V
32 0 V
31 0 V
32 -1 V
32 0 V
31 0 V
32 0 V
31 0 V
32 0 V
3.000 UL
LT1
450 1662 M
31 0 V
32 0 V
31 0 V
31 0 V
32 0 V
31 -1 V
31 -1 V
32 -1 V
31 -3 V
31 -3 V
32 -3 V
31 -5 V
31 -5 V
32 -5 V
31 -4 V
31 -3 V
31 -3 V
32 -1 V
31 -1 V
31 -1 V
32 0 V
31 -1 V
31 0 V
32 0 V
31 0 V
31 0 V
32 0 V
31 -1 V
31 0 V
32 0 V
31 0 V
31 0 V
32 0 V
31 0 V
31 0 V
32 0 V
31 0 V
31 0 V
32 0 V
31 0 V
31 0 V
32 0 V
31 0 V
31 0 V
31 0 V
32 0 V
31 0 V
31 0 V
32 0 V
31 0 V
31 0 V
32 0 V
31 0 V
31 0 V
32 0 V
31 0 V
31 0 V
32 0 V
31 0 V
3.000 UL
LT3
450 1674 M
31 0 V
32 -1 V
31 -1 V
31 -2 V
32 -3 V
31 -5 V
31 -8 V
32 -13 V
31 -19 V
31 -30 V
32 -44 V
31 -64 V
31 -87 V
32 -112 V
31 -129 V
31 -136 V
982 893 L
32 -108 V
31 -84 V
31 -60 V
32 -42 V
31 -28 V
31 -19 V
32 -12 V
31 -7 V
31 -5 V
32 -3 V
31 -2 V
31 -2 V
32 -1 V
31 0 V
31 -1 V
32 0 V
31 0 V
31 0 V
32 -1 V
31 0 V
31 0 V
32 0 V
31 0 V
31 0 V
32 -1 V
31 0 V
31 0 V
31 0 V
32 0 V
31 0 V
31 0 V
32 -1 V
31 0 V
31 0 V
32 0 V
31 0 V
31 0 V
32 0 V
31 0 V
31 0 V
32 0 V
31 0 V
1.000 UL
LT4
550 413 M
263 0 V
450 1668 M
31 0 V
32 0 V
31 0 V
31 -1 V
32 0 V
31 -1 V
31 -2 V
32 -3 V
31 -4 V
31 -7 V
32 -9 V
31 -13 V
31 -16 V
32 -20 V
31 -20 V
31 -19 V
31 -16 V
32 -13 V
31 -9 V
31 -7 V
32 -4 V
31 -3 V
31 -2 V
32 -1 V
31 -1 V
31 0 V
32 -1 V
31 0 V
31 0 V
32 -1 V
31 0 V
31 0 V
32 0 V
31 0 V
31 0 V
32 0 V
31 0 V
31 -1 V
32 0 V
31 0 V
31 0 V
32 0 V
31 0 V
31 0 V
31 0 V
32 0 V
31 0 V
31 0 V
32 0 V
31 0 V
31 -1 V
32 0 V
31 0 V
31 0 V
32 0 V
31 0 V
31 0 V
32 0 V
31 0 V
1.000 UL
LT6
450 1668 M
32 0 V
33 -1 V
32 -1 V
33 -1 V
32 -3 V
33 -4 V
32 -6 V
32 -9 V
33 -15 V
32 -23 V
33 -32 V
32 -44 V
32 -56 V
33 -61 V
32 -62 V
33 -55 V
32 -43 V
33 -33 V
32 -22 V
32 -14 V
31 -10 V
32 -6 V
31 -4 V
32 -3 V
32 -1 V
31 -1 V
32 -1 V
31 -1 V
32 0 V
31 0 V
32 0 V
32 0 V
31 -1 V
32 0 V
31 0 V
32 0 V
32 0 V
31 0 V
32 0 V
31 0 V
32 0 V
32 0 V
31 -1 V
32 0 V
31 0 V
32 0 V
32 0 V
31 0 V
32 0 V
31 0 V
32 0 V
31 0 V
32 0 V
32 0 V
31 0 V
32 0 V
31 0 V
32 0 V
1.000 UL
LT7
450 1668 M
31 -1 V
32 -1 V
31 -2 V
31 -2 V
32 -4 V
31 -7 V
31 -9 V
32 -16 V
31 -24 V
31 -36 V
32 -52 V
31 -72 V
31 -92 V
32 -107 V
31 -114 V
31 -106 V
31 -90 V
32 -70 V
31 -51 V
31 -34 V
32 -24 V
31 -15 V
31 -9 V
32 -6 V
31 -4 V
31 -3 V
32 -2 V
31 -1 V
31 0 V
32 -1 V
31 0 V
31 0 V
32 0 V
31 -1 V
31 0 V
32 0 V
31 0 V
31 0 V
32 0 V
31 0 V
31 0 V
32 0 V
31 -96 V
31 -210 V
9 -106 V
stroke
grestore
end
showpage
}}%
\put(863,413){\makebox(0,0)[l]{$a_B=-0.01$}}%
\put(1374,822){\makebox(0,0)[l]{\scriptsize$a_B=-0.0559$}}%
\put(1374,1219){\makebox(0,0)[l]{\scriptsize$a_B=-0.03$}}%
\put(1960,572){\makebox(0,0)[l]{\scriptsize$N=30$}}%
\put(1960,1585){\makebox(0,0)[l]{\scriptsize$N=10$}}%
\put(1682,1888){\makebox(0,0)[l]{\scriptsize$\lambda_3$}}%
\put(1682,1721){\makebox(0,0)[l]{\scriptsize$\lambda_2$}}%
\put(597,1888){\makebox(0,0)[l]{\scriptsize$\lambda_1$}}%
\put(1374,2226){\makebox(0,0){ }}%
\put(1374,50){\makebox(0,0){$\rho/b$}}%
\put(100,1188){%
\special{ps: gsave currentpoint currentpoint translate
270 rotate neg exch neg exch translate}%
\makebox(0,0)[b]{\shortstack{$\lambda$}}%
\special{ps: currentpoint grestore moveto}%
}%
\put(2298,200){\makebox(0,0){$10^{5}$}}%
\put(1990,200){\makebox(0,0){$10^{4}$}}%
\put(1682,200){\makebox(0,0){$10^{3}$}}%
\put(1374,200){\makebox(0,0){$10^{2}$}}%
\put(1066,200){\makebox(0,0){$10^{1}$}}%
\put(758,200){\makebox(0,0){$10^{0}$}}%
\put(450,200){\makebox(0,0){$10^{-1}$}}%
\put(400,2076){\makebox(0,0)[r]{200}}%
\put(400,1867){\makebox(0,0)[r]{100}}%
\put(400,1658){\makebox(0,0)[r]{0}}%
\put(400,1449){\makebox(0,0)[r]{-100}}%
\put(400,1240){\makebox(0,0)[r]{-200}}%
\put(400,1031){\makebox(0,0)[r]{-300}}%
\put(400,822){\makebox(0,0)[r]{-400}}%
\put(400,613){\makebox(0,0)[r]{-500}}%
\put(400,404){\makebox(0,0)[r]{-600}}%
\end{picture}%
\endgroup
 